\documentclass[aps,prd,nofootinbib]{revtex4}
\usepackage[dvips]{graphicx}
\usepackage{bm,latexsym,amsmath,amssymb,amsfonts}

\begin{document}

\title{Static black hole uniqueness and Penrose inequality}

\author{${}^{(a)}$Ryosuke Mizuno, 
${}^{(a),(b)}$Seiju Ohashi, ${}^{(a)}$Tetsuya Shiromizu}

\affiliation{${}^{(a)}$Department of Physics, Kyoto University, 
Kyoto 606-8502, Japan}
\affiliation{${}^{(b)}$Department of Physics, Tokyo Institute of Technology,
Tokyo 152-8551, Japan}

\begin{abstract}
Under certain conditions, we give a new way to prove the uniqueness of 
static black hole in higher dimensional asymptotically flat spacetimes. 
In the proof, the Penrose inequality plays a key role in higher dimensions 
as well as four dimensions. 
\end{abstract}


\maketitle

\section{Introduction}

Although general relativity plays central role in cosmology and astrophysics, 
there are still fundamental open problems. Among them, so called cosmic censorship 
conjecture is longstanding issue initiated by Penrose \cite{Penrose:1969pc}. 
Related to this, Penrose 
proposed an inequality which is named by Penrose inequality. If counter examples 
against this inequality exist, it is unlikely that cosmic censorship conjecture holds. 
The Penrose inequality is the conjecture that the area of the black hole horizon 
in any spacetimes is equal or less then the area of the Schwarzschild black hole that contains 
the same ADM mass \cite{Pen}. When the equality holds, the spacetime will be the Schwarzschild 
spacetime. This conjecture also implies that the Schwarzschild black hole has a 
maximum entropy in the context of the black hole thermodynamics. Therefore, it is 
important to confirm this conjecture in the details from several points of view. 
See Ref. \cite{Mars} for recent status of the Penrose inequality. 

At first, after some pioneering works \cite{Geroch, Jang}, the Penrose inequality 
was proven for a single black hole on three dimensional time-symmetric initial data 
by Huisken and Ilmanen using the inverse mean curvature flow \cite{IMC}. 
Then Bray proved the Penrose inequality for multi black holes using the conformal flow \cite{Bray} and also his proof was 
extended to higher dimensions less than eight space dimensions later \cite{Bray2}. However, 
the Penrose inequality on the arbitrary time slice has not been proven. While such a development, 
in Ref. \cite{Chrusciel:2000az}, the relation between the Penrose inequality and 
the uniqueness theorem for black hole with a negative cosmological constant 
in four dimensions has been discussed. That is, if a certain inequality like 
the Penrose inequality holds, one can show the uniqueness of black hole spacetimes 
with a negative cosmological constant. {\it It is reminded that 
this observation is plugged into the old proof of static black hole by Israel} \cite{Israel}, {\it although 
he did not mention it explicitly} (this is because the Penrose inequality was proposed after 
Ref. \cite{Israel} appears!). 
 Inspired by these observations, we shall discuss the same issue in asymptotically flat 
and higher dimensional spacetimes. 

In static and asymptotically flat spacetimes, the uniqueness theorem of higher 
dimensional black hole spacetimes has been proven \cite{GIS}. Therein the key 
tools were the conformal transformation and the positive energy theorem as 
well as four dimensional cases \cite{BM}. And, using the same 
tool with the uniqueness theorem, the Riemannian Penrose inequality was also 
proven on time-symmetric initial data \cite{Bray2}. Apart from the 
Israel-type proof, at first glance, the Penrose inequality does not appear in the 
proof of the uniqueness. Since same tool 
was used in both proofs, however, we would guess the presence of the 
deep relation between them. To confirm this, using the Penrose-type inequality, 
we try to prove the uniqueness theorem in an independent way. 

The rest of this paper is organized as follows. In Sec. II, following 
the Israel-type proof \cite{Israel}, we 
first try to prove the uniqueness of static black hole in higher dimensions and review 
the four dimensional argument briefly. Then we realize that the 
straightforward extension of the proof from four to higher dimensions 
is impossible and the role of the Penrose inequality is 
important in the proof. That is, we must prove the Penrose-type inequality 
for the proof of the uniqueness. In Sec. III, we will present a new way 
to prove the uniqueness through proving the Penrose-type inequality with 
certain conditions. Finally we summarize our work and have a comment. 
We also discuss the remaining issues left for future studies.

\section{Trial work for uniqueness}

In this section, we shall try to prove the uniqueness of static black hole spacetimes 
following the idea of Israel's original version for four dimensions \cite{Israel}. In 
addition, we  point out that the Penrose inequality is tacitly used in the proof in 
four dimensions and realize that the straightforward extension is not possible. Then 
we make the problem clear. 

We consider the $n$-dimensional vacuum spacetime satisfying $R_{ab}=0$ in higher dimensions 
and assume that the horizon is topologically sphere. 
The metric of a static spacetime is written as 
\begin{equation}
ds^2=-V^2(x^i)dt^2+g_{ij}(x^k)dx^idx^j,
\end{equation}
where the indices $i,j,k, \cdots $ stand for the spatial components. 
The event horizon $H$ is located at $V=0$. 
The Einstein equations become
\begin{eqnarray}
D^2V=0
\end{eqnarray}
and
\begin{eqnarray}
{}^{(n-1)}R_{ij}=\frac{1}{V}D_i D_j V,
\end{eqnarray}
where $D_i$ is the covariant derivative with respect to $g_{ij}$ and ${}^{(n-1)}R_{ij}$ is 
the Ricci tensor of $g_{ij}$. 
Since $V$ is a harmonic function, we can employ it as a kind of ``radial" coordinate 
\begin{equation}
ds^2=-V^2dt^2+\rho^2dV^2+h_{AB}dx^Adx^B,
\end{equation}
where $\rho=(D^iVD_iV)^{-1/2}$ which follows from the definition. 
The indices $A,B \cdots$ stand for the orthogonal component to $V={\rm const.}$ 
surfaces. 

From the vacuum Einstein equations we have the following relations
%
\begin{eqnarray}
{}^{(n-2)}R=\frac{2}{V\rho}k+k^2-k_{ab}k^{ab}\label{decomp}
\end{eqnarray}
%
and
%
\begin{eqnarray}
\partial_V k=\frac{k}{V}-\rho k_{ab}k^{ab}-{\cal D}^2 \rho, 
\end{eqnarray}
%
where $k_{ab}$ and $k$ is the extrinsic curvature and the mean curvature of $V=$const. surfaces, respectively, and 
$\cal D$ is the covariant derivative with respect to $h_{ab}$. 

In addition, we also have the following three equations
%
\begin{eqnarray}
\partial_V \Biggl( \frac{{\sqrt {h}}}{\rho}\Biggr)=0, \label{eq1}
\end{eqnarray}
%
%
\begin{eqnarray}
\partial_V \Biggl( \frac{{\sqrt {h}}}{\rho^{\frac{n-3}{n-2}}} \frac{k}{V}\Biggr)=
-\frac{{\sqrt {h}}}{V} 
\Biggl[ (n-2) {\cal D}^2 \rho^{\frac{1}{n-2}}+\frac{1}{\rho^{\frac{2n-5}{n-2}}}\Bigl( \rho^2 \tilde k_{ab} \tilde k^{ab}
+ \frac{n-3}{n-2} ({\cal D} \rho)^2 \Bigr) 
\Biggr] \label{eq2}
\end{eqnarray}
%
and
%
\begin{eqnarray}
\partial_V \Biggl[ \frac{{\sqrt {h}}}{\rho} \Biggl( \frac{n-3}{n-2}kV+\frac{2}{\rho} \Biggr) \Biggr]=
-{\sqrt {h}}V \Biggl[ 
\frac{1}{n-2}{}^{(n-2)}R+\rho^{-2}\Bigl(\frac{n-3}{n-2}({\cal D} \rho)^2 +\rho^2 \tilde k_{ab} \tilde k^{ab}\Bigr)
+\frac{n-3}{n-2}{\cal D}^2 \ln \rho \Biggr], \label{eq3}
\end{eqnarray}
%
where $\tilde k_{ab}=k_{ab}-\frac{1}{n-2}h_{ab}k$ and $h=\det(h_{AB})$. 

We focus on asymptotically flat spacetimes. It is easy to check that 
asymptotic behaviors of some geometrical quantities at spatial 
infinity are given by 
%
\begin{eqnarray}
V \simeq 1-\frac{m}{r^{n-3}},
\end{eqnarray}
%
%
\begin{eqnarray}
k \simeq \frac{n-2}{r}
\end{eqnarray}
%
and
%
\begin{eqnarray}
\rho \simeq \frac{1}{n-3}\frac{r^{n-2}}{m},
\end{eqnarray}
%
where $m$ is the ADM mass of the spacetime.
To address the regularity at the event horizon $H$, we compute $R_{abcd}R^{abcd}$ and the 
result is given by 
%
\begin{eqnarray}
R_{abcd}R^{abcd}={}^{(n-1)}R_{ijkl}{}^{(n-1)}R^{ijkl}+\frac{4}{V^2\rho^2}
\Biggl(k_{ab}k^{ab}+\frac{2}{\rho^2}({\cal D} \rho)^2+\frac{1}{\rho^4}(\partial_V \rho)^2\Biggr).
\end{eqnarray}
%
Then the regularity at the horizon implies
%
\begin{eqnarray}
k_{ab}|_H={\cal D}_a \rho|_H=0.
\end{eqnarray}
%
In addition, Eq. (\ref{decomp}) gives us 
%
\begin{eqnarray}
{}^{(n-2)}R|_H=2(k/V\rho)|_H. \label{ricci}
\end{eqnarray}
%

Let us take the volume integral of Eq. (\ref{eq1}) in the $t=$const. hypersurface. Then 
we have 
%
\begin{eqnarray}
(n-3)m \Omega_{n-2}=\frac{S_H}{\rho_H}, \label{equality1}
\end{eqnarray}
%
where $S_H$ is the area of the event horizon, $\Omega_{n-2}$ is 
the area of the unit $(n-2)$-sphere and $\rho_H=\rho|_{H}$ . 
From Eq. (\ref{eq2}) with Eqs. (\ref{equality1}) and (\ref{ricci}), we obtain
%
\begin{eqnarray}
(n-2)[(n-3)m]^{(n-3)/(n-2)} \Omega_{n-2} \leq \frac{1}{2}\rho_H^{1/(n-2)}
\int_{S_H}dS{}^{(n-2)}R. \label{ineq1}
\end{eqnarray}
%
From Eq. (\ref{eq3}), we see 
%
\begin{eqnarray}
4S_H/\rho_H^2 \geq \frac{2}{n-2}\int_0^1dVV \int_{S_V}dS{}^{(n-2)}R, \label{ineq2}
\end{eqnarray}
%
where $S_V$ is a $V=$const. surface. 

To proceed the proof, it is better to review Israel's argument in 
four dimensions ($n=4$). In this case, Gauss-Bonnet theorem tells us 
that $\int_{S_H}{}^{(2)}R=
\int_{S_V}{}^{(2)}R=8\pi$ holds. Therefore, two inequalities become
%
\begin{eqnarray}
2m^{1/2} \leq \rho_H^{1/2}
\end{eqnarray}
%
and
%
\begin{eqnarray}
S_H \geq \pi \rho_H^{2},
\end{eqnarray}
%
respectively. Using Eq. (\ref{equality1}), the first one becomes 
the reverse Penrose inequality
%
\begin{eqnarray}
4\pi (2m)^2 \leq S_H. \label{iPen}
\end{eqnarray}
%
On the other hand, the second one is just the Penrose inequality
%
\begin{eqnarray}
4\pi (2m)^2 \geq S_H.
\end{eqnarray}
%
Then we can show the equality, $4\pi (2m)^2 = S_H$, and this implies  
that $\tilde k_{ab}={\cal D}_a \rho=0$ holds. Thus, it is turned out that 
the spacetime must be spherical symmetric, that is, the Schwarzschild 
spacetime.

In the above, we can see that the Penrose inequality {\it explicitly} appears in 
Israel's proof. However, it is the 
story in four dimensions. Our current end is to address if the same argument can work. 
However we immediately realize that we cannot use the inequality of Eq. (\ref{ineq2}) which 
may provide us the Penrose inequality. This is because we cannot evaluate $\int_{S_V}{}^{(n-2)}RdS$
 which is not a topological invariant in higher dimensions. Thus we need a new ingredient for proving 
the uniqueness. 

Here we note that the uniqueness of static black hole has been proven 
using a different way \cite{GIS}. But we remind that 
the proof of the uniqueness itself is not our current end. What we want to see is the 
direct relation between the Penrose inequality and the uniqueness. To see this, we shall follow 
Israel's way with slight modification of the argument around the inequality of Eq. (\ref{ineq2}). 

\section{Penrose-like inequality}

Let us introduce the following dimensionless geometrical quantity 
%
\begin{eqnarray}
Y_H:=\frac{\int_{H}dS {}^{(n-2)}R}{S_H^{(n-4)/(n-2)}}
\end{eqnarray}
%
which is a mimic of the Yamabe invariant. From Eq. (\ref{ineq1}), we see that $Y_H$ is 
positive (see Ref. \cite{Galloway} for general argument). 
If the horizon is spherical symmetric, $Y_H$ becomes 
%
\begin{eqnarray}
Y_H^0=(n-2)(n-3)\Omega_{n-2}^{2/(n-2)}. 
\end{eqnarray}
%
For the convenience, it is nice to normalize $Y_H$ by $Y_H^0$ as 
%
\begin{eqnarray}
y_H:=\frac{Y_H}{Y_H^0}. \label{const}
\end{eqnarray}
%
Using $y_H$ and Eq. (\ref{equality1}), the inequality of Eq. (\ref{ineq1}) is rearranged as 
%
\begin{eqnarray}
(2y_H^{-1}m)^{\frac{n-2}{n-3}}\Omega_{n-2} \leq S_H. \label{ipi}
\end{eqnarray}
%
This corresponds to Eq. (\ref{iPen}) in four dimensions (note that $y_H=1$ holds in 
four dimensions due to the Gauss-Bonnet theorem). 
If we can show the inverse version of the above inequality, 
we can show the spherical symmetry of the spacetime. To show this, 
we shall employ a primitive version of the inverse 
mean curvature flow introduced by Geroch \cite{Geroch} 
(see also Ref. \cite{Gibbons} for a trial work in higher dimensions). 
Here we note that the Penrose inequality 
$(2m)^{\frac{n-2}{n-3}}\Omega_{n-2} \geq S_H$
was proven \cite{Bray2}. However, we cannot use this for the current 
purpose due to the luck of $y_H$-dependence. 

Without loss of generality, we can write down the unit normal vector of 
$(n-2)$ surfaces as $\varphi^{-1}\partial_z$ and we set $z=0$ surface to 
be the minimal surface (that is the event horizon of the spacetime). Note 
that $z=$const. 
level surfaces are different from $V=$const. level surfaces. First one may propose the 
following quasi-local mass as 
%
\begin{eqnarray}
m(z)= \frac{S^{\frac{1}{n-2}}}{2(n-2)(n-3)\Omega_{n-2}^{\frac{n-1}{n-2}}}
\int_{z={\rm const}} dS \Biggl({}^{(n-2)}\bar R -\frac{n-3}{n-2}\bar k^2 \Biggr), \label{difm}
\end{eqnarray}
%
where $\bar k_{ab}$ and ${}^{(n-2)}\bar R$ are the extrinsic curvature and 
the intrinsic curvature of $z=$const. surfaces. In four dimensions ($n=4$), it 
becomes the Hawking quasi-local mass \cite{Hawking}. We can check that it agrees 
with the ADM mass at the spatial infinity (it is supposed to correspond to $z=\infty$), 
that is, $m(\infty)=m$. 

Now we define the following function 
%
\begin{eqnarray}
f(z)=\int_{z={\rm const}}dS \Biggl({}^{(n-2)}\bar R -\frac{n-3}{n-2}\bar k^2 \Biggr).
\end{eqnarray}
%
Choosing the mean curvature so that $\bar k \varphi=1$ is satisfied, 
the first variation of $f(z)$ becomes 
%
\begin{eqnarray}
\frac{df(z)}{dz}=-\frac{1}{n-2}f(z)
-2\int {}^{(n-2)}\tilde{\bar R}_{ab} \tilde {\bar k}^{ab}\varphi dS
+\frac{n-3}{n-2}\int dS \Biggl[2\varphi^{-2}({\cal D} \varphi)^2 +
\tilde {\bar k}_{ab}\tilde {\bar k}^{ab}+{}^{(n-1)}R \Biggl], \label{df}
\end{eqnarray}
%
where $\tilde{\bar R}_{ab}$ and $\tilde {\bar k}^{ab}$ is the trace free part 
of ${}^{(n-2)}{\bar R}_{ab}$ and $\bar k_{ab}$, respectively,
and $\bar {\cal D}_a$ is the covariant derivative with respect to the metric of 
$z=$const. surfaces. Using the vacuum Einstein equations, we can show that 
${}^{(n-1)}R=0$ holds on time-symmetric initial data. Then we see the inequality 
%
\begin{eqnarray}
\frac{df(z)}{dz} \geq -\frac{1}{n-2}f(z)
-2\int dS {}^{(n-2)}\tilde{\bar R}_{ab} \tilde {\bar k}^{ab}\varphi 
\end{eqnarray}
%
holds and 
%
\begin{eqnarray}
e^{\frac{1}{n-2}z}f(z) \geq 
f(0)-2\int^z_0 dz' e^{\frac{1}{n-2}z'} 
\int_{z'={\rm const}} dS{}^{(n-2)}\tilde{\bar R}_{ab} \tilde {\bar k}^{ab}\varphi.
\end{eqnarray}
%
From $\bar k\varphi=1$, we also have 
%
\begin{eqnarray}
\frac{dS(z)}{dz}=\int dS \bar k\varphi =S(z)
\end{eqnarray}
%
and then $S(z)=e^z S_H$. Using Eq. (\ref{difm}), finally we obtain 
%
\begin{eqnarray}
m \geq \frac{1}{2}\Bigl(\frac{S_H}{\Omega_{n-2}}\Bigr)^{\frac{n-3}{n-2}}y_H
-\frac{S_H^{\frac{1}{n-2}}}{(n-2)(n-3)\Omega_{n-2}^{\frac{n-1}{n-2}}}
\int^\infty_0 dz' e^{\frac{1}{n-2}z'} 
\int_{z'={\rm const}} dS{}^{(n-2)}\tilde{\bar R}_{ab} \tilde {\bar k}^{ab}\varphi.
\end{eqnarray}
%
Here we suppose the following condition
%
\begin{eqnarray}
\int^\infty_0 dz' e^{\frac{1}{n-2}z'}
\int_{z'={\rm const}} dS{}^{(n-2)}\tilde{\bar R}_{ab} \tilde {\bar k}^{ab}\varphi \leq 0. \label{cond}
\end{eqnarray}
%
For instance, one sees that the above is satisfied if the metric is Einstein, that is,
%
\begin{eqnarray}
{}^{(n-2)}\tilde{\bar R}_{ab}=0 \label{cond2}.
\end{eqnarray}
%
This is also rather strong condition, but still covers a wide class of manifolds, that is, 
manifold corresponding to extrema of $y_H$ for the variation of the metric $h_{ab}$.
In particular, Eq. (\ref{cond2}) holds in a round $(n-2)$-sphere.
Then we obtain a mimic of the Penrose inequality
%
\begin{eqnarray}
(2my_H^{-1})^{\frac{n-2}{n-3}}\Omega_{n-2} \geq S_H. \label{mpi}
\end{eqnarray}
%
Together with Eq. (\ref{ipi}), we see the equality 
should hold, that is, 
%
\begin{eqnarray}
(2my_H^{-1})^{\frac{n-2}{n-3}}\Omega_{n-2} = S_H. 
\end{eqnarray}
%
This implies 
%
\begin{eqnarray}
\tilde{k}_{ab}={\cal D}_a \rho=0.
\end{eqnarray}
%
This means that the spacetime is spherically symmetric. In the same way as 
four dimensional cases, it is easy to show that spherical symmetric 
vacuum spacetimes must be the higher dimensional Schwarzschild spacetime. 

We could show that, under the condition of Eq. (\ref{cond}) or (\ref{cond2}), 
the uniqueness of static black hole spacetimes via the Penrose-like 
inequality of Eq. (\ref{mpi}) holds. Thus the Penrose-like inequality plays a 
key role in our argument. 

\section{summary and discussion}

Let us summarize our current work. In this paper, we presented a new way to prove the 
uniqueness of static black hole in higher dimensional asymptotically flat 
spacetimes. Therein we saw the importance of the Penrose-like 
inequality in the new proof of the uniqueness theorem of black holes in higher dimensions. 
In asymptotically flat static vacuum spacetimes, we have the reverse Penrose-like 
inequality (Eq. (\ref{ipi})), then the Penrose-like inequality (Eq. (\ref{mpi})) implies the 
uniqueness theorem. 

Here we have several comments. Under the assumption of ${}^{(n-2)}\tilde {\bar R}_{ab}=0$ 
on the horizon, 
it is known that $y_H \leq 1$ holds (for example, see Proposition 1.4 in \cite{Schoen} ). 
Then Eq. (\ref{ipi}) becomes $(2m)^{(n-2)/(n-3)}\Omega_{n-2} \leq S_H$. On the other hand, 
the Riemannian Penrose inequality, $(2m)^{(n-2)/(n-3)}\Omega_{n-2} \geq S_H$, 
has been proven in Ref. \cite{Bray2} (it works for higher dimensions less than eight). 
Thus, we can see that the equality holds and then it implies the 
spherical symmetry. This gives us an alternative way to prove the uniqueness. 
Note that the positive mass theorem is used therein. Since the positive 
mass theorem is also used in the direct proof of the uniqueness, yet, 
without the condition of ${}^{(n-2)}\tilde {\bar R}_{ab}=0$, 
this argument is not so clever as the proof of the uniqueness. Nevertheless, this also indicates us the presence of the 
deep relation between the uniqueness theorem and the Penrose inequality. 

Since the condition of Eq. (\ref{cond}) or (\ref{cond2}) are rather strong. Therefore 
it is better to remove it. To do so, we may employ other foliations with the gauge 
ambiguity, which is different from the inverse mean curvature flow. The gauge ambiguity 
will be used to drop the trouble term to show the Penrose-type inequality. 

In the derivation of the Penrose-like inequality (Eq. (\ref{mpi})), we used ${}^{(n-1)}R=0$ 
which comes from the vacuum Einstein equation. However, if ${}^{(n-1)}R \geq 0$ is satisfied, 
which corresponds to the dominant energy condition in a time-symmetric initial data,  
the third term in the right-hand side of Eq. (\ref{df}) is shown to be 
non-negative. Thus, the Penrose-like inequality (Eq. (\ref{mpi})) still holds on a 
time-symmetric data that satisfies the dominant energy condition and Eq. (\ref{cond}). 

The inequality of Eq. (\ref{mpi}) contains $Y_H$ that is a mimic of the 
Yamabe invariant on the horizon. Here we remember that 
Penrose-like inequality for higher dimensions proven in Ref. 
\cite{Herzlich2} also depends on the Yamabe invariant on the event horizon 
in non-trivial way. In a numerical analysis, one can confirm that the Penrose 
inequality holds \cite{YS}. But, it is nice to see the dependence of the 
Yamabe invariant or so.


\acknowledgments
We are grateful to Sumio Yamada for useful discussions. 
TS is supported by Grant-Aid for Scientific Research from Ministry of 
Education, Science, Sports and Culture of Japan (Nos.~20540258 and 19GS0219), 
the Japan-U.K., Japan-France and Japan-India Research Cooperative Programs.




\end{document}